\documentclass[aps,prl,twocolumn,groupedaddress,amsmath,amsfonts,amssymb]{revtex4}

\usepackage{graphicx}
\usepackage{bm}
\usepackage{color}
\usepackage{nicefrac}
\usepackage{amsmath}
\usepackage{verbatim}

\newcommand{\bra}[1]{\langle #1|}
\newcommand{\ket}[1]{|#1\rangle}
\newcommand{\braket}[2]{\langle #1|#2\rangle}
\newcommand{\mm}[1]{\mathrm{#1}}
\newcommand{\el}[2]{^{#2}\mathrm{#1}}

\newcommand{\dplus}{\delta_{jj'+1}}
\newcommand{\dminus}{\delta_{jj'-1}}
\newcommand{\NA}{N^{AB}_{nn'}}
\newcommand{\NB}{N^{AB}_{n'n}}

\newcommand{\dr}{\mathrm{d}r}
\newcommand{\HRA}{(H_R)^{\uparrow\downarrow}_{nn'}}

\newcommand{\Ze}{\frac{1}{2}g\mu_{B}B}
\newcommand{\unit}[1]{\ensuremath{\, \mathrm{#1}}}
\newcommand{\del}{\partial}

\renewcommand{\vec}[1]{\mathbf{#1}}

\begin{document}

\title{Effective time-reversal symmetry breaking in the spin relaxation \\in a graphene quantum dot}

\author{P.\ R.\ Struck}

\author{Guido Burkard}

\affiliation{Department of Physics, University of Konstanz, D-78457 Konstanz, Germany}


\date{\today}

\begin{abstract}

We study the relaxation of a single electron spin in a circular gate-tunbable quantum dot in gapped graphene.
Direct coupling of the electron spin to out-of-plane phonons via the intrinsic spin-orbit coupling
leads to a relaxation time $T_1$ which is independent of the B-field at low fields. 
We also find that Rashba spin-orbit induced admixture of opposite spin states in combination with the
emission of in-plane phonons provides various further relaxation channels via deformation potential and bond-length change.
In the absence of valley mixing, spin relaxation takes place within each valley separately and thus
time-reversal symmetry is effectively broken, thus inhibiting the van Vleck cancellation 
at $B=0$ known from GaAs quantum dots.  
Both the absence of the van Vleck cancellation as well as the out-of-plane phonons
lead to a behavior of the spin relaxation rate at low magnetic fields which is markedly
different from the known results for GaAs.
For low B-fields, we find that the rate is constant in $B$ and then crosses
over to $\propto B^2$ or $\propto B^4$ at higher fields.
\end{abstract}

\maketitle

\textit{Introduction---}The electronic spin degree of freedom is under intense investigation as a possible implementation of a qubit \cite{loss_divicenzo_proposaL_QD}.
While the feasibility of all required operations has been experimentally demonstrated for GaAs quantum dots (QDs) \cite{hanson}, the decoherence caused by
the surrounding nuclear spins in the host material remains challenging.
Regarding the use of the electron spin as a qubit in quantum computation devices, spin decoherence and relaxation are limiting factors. In general, a necessary condition for a working qubit is that the time required to perform an operation is significantly shorter than the decoherence and relaxation times.  Motivated by this, the implementation of qubits in QDs in graphene has been proposed \cite{trauzettel}. 
Graphene consisting of natural carbon comprises 99\% of the carbon isotope $\el{C}{12}$ which has no nuclear spin,  hence the hyperfine interaction is expected to play only a minor role.  Furthermore, spin-orbit interaction (SOI) in graphene is expected to be relatively weak and therefore long decoherence times are expected.  However, for spins localized in QDs in 
carbon nanotubes, SOI has turned out to be unexpectedly
strong \cite{kuemmeth,churchill} due to curvature-induced effects.
It has also been shown theoretically that van Hove singluarities in the
phonon density of states in one dimension can lead to strong
variations in the spin relaxation rate \cite{bulaev}. 
It is therefore important to investigate the spin relaxation time in graphene QDs.
The form of the SOI in graphene, both intrinsic and Rashba type, 
is known \cite{kane}, and there are several works 
on its strength depending on various parameters such as curvature or electric field
\cite{huertas06,min}.  There have also been  experimental \cite{tombros} 
and theoretical \cite{huertas09} studies on spin relaxation of extended states in graphene.

In this paper we determine theoretically the spin relaxation time $T_1$ for an electron confined to a 
circular QD in gapped graphene as a function of the external magnetic field $\vec{B}$.
It has been predicted previously that such QDs can be formed with electrostatic confinement in either single-layer graphene with a substrate-induced band gap or bilayer graphene with an electrically controlled gap \cite{circular_QD}.  At $\vec{B}=0$, the states in these
QDs have a two-fold valley degeneracy which can be lifted in a perpendicular magnetic field. 
\begin{figure}
	\includegraphics[width=0.48 \textwidth]{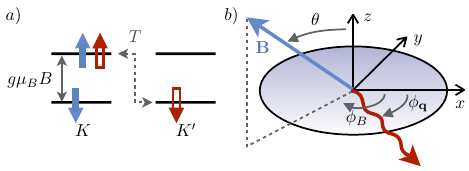}
	\caption{(Color online) a) The two states of a \textit{spin qubit}
(blue solid arrows) reside in the same valley, as opposed to a \textit{Kramers qubit} 
(empty red arrows), formed by a Kramers pair related by time-reversal symmetry ($T$). 
While in single-valley semiconductors such as GaAs these two cases are identical, in graphene the
Kramers qubit involves states in different valleys ($K$ and $K'$).  
b) The $B$-field orientation is given with 
spherical coordinates $\theta$ and $\phi_B$ relative to the normal to the graphene plane.  
The propagation direction of the emitted phonon (red wavy arrow) is 
described by the angle $\phi_q$.}
	\label{fig:figure1}
\end{figure}
Being a centro-symmetric crystal, phonons in graphene do not couple piezo-electrically, thus
leaving three possible electron-phonon coupling (EPC) mechanisms: deformation potential, bond length
change, and direct spin-phonon coupling.  From these EPC mechanisms, we derive 
two spin relaxation mechanisms.
One such mechanism involves the admixture of states of
opposite spin and excited orbitals into the dot eigenstates due to SOI, 
in combination with energy relaxation via phonon emission \cite{van_vleck,nazarov}.
It turns out that to lowest order in the EPC, this only involves in-plane phonons coupled via
the deformation potential and bond-length change.
The second mechanism directly couples the spin to out-of-plane phonons via curvature induced SOI.
For comparison, in a parabolic GaAs QD, a strong dependence $\propto B^5$ has been predicted for both mechanisms \cite{nazarov}. Relaxation times in the millisecond range at a field $B=1\,{\rm T}$ have been predicted and even longer $T_1$ exceeding one second have been experimentally verified \cite{zumbuehl}.
The prediction for graphene QDs looks markedly different because of
the absence of the van Vleck cancellation for spin qubits in a single valley 
as opposed to ``Kramers qubits'' (see Fig.~\ref{fig:figure1}a), as well as
the absence of piezo phonons and the two-dimensional nature of phonons.

\textit{Model---}To study spin relaxation in a circular and gate-tunable QD in single-layer graphene, we assume the host graphene layer to be sufficiently large to ensure that the edges do not induce inter-valley mixing. 
The QD can be described with the Hamiltonian \cite{circular_QD},
\begin{equation}
\label{H0}
H_{0}=v_{F}(\vec{p}+e\vec{A}_{\perp})\cdot\vec{\boldsymbol\sigma}
+\frac{1}{2}g\mu_{B}\vec{B}\cdot\vec{s}
+U(r)
+\tau\Delta\sigma_{z},
\end{equation}
where the first term is the well-known Dirac Hamiltonian for graphene \cite{castro_neto} 
in the presence of a vector potential $\vec{A}_{\perp}$ with $\vec{B}_{\perp}=\nabla\times\vec{A}_{\perp}=(0,0,B\cos\theta)$ being the perpendicular component of an arbitrarily oriented B-field
(Fig.~\ref{fig:figure1}b).  
The second and third
terms describe the Zeeman coupling of the electron spin to the total B-field
and the smooth and circularly symmetric confinement potential $U(r)$.
The last term opens a band gap
$2\Delta$ which can e.g.\ arise due to the influence of the substrate 
\cite{giovannetti,zhou}.  Here, $\tau=\pm 1$ denote the $K$ and $K'$ valleys.
In the absence of valley-scattering, we can restrict ourselves to a single
valley, e.g. $\tau=+1$.  Weak inter-valley coupling can arise from atomic 
defects or boundaries \cite{castro_neto}, or from the hyperfine interaction with the remaining $^{13}$C atoms \cite{andras}.

The eigenstates $\ket{n,s}^{(0)}$ of $H_0$ in Eq.~\eqref{H0} with 
energy $E_{n}+s g \mu_{B}B/2$
are simultaneously eigenstates of the total angular momentum $j\in\mathbb{Z}+\frac{1}{2}$, 
i.e.\ orbital quantum number and pseudo-spin, 
with spinor wavefunctions (in the $K$ valley),
\begin{equation}
	\braket{r,\phi}{n;s} =
	\psi_{n}(r,\phi) = 
	e^{i(j-1/2)\phi}
	\begin{pmatrix}\chi_{A}^{j,\nu}(r) \\ 
		\chi_{B}^{j,\nu}(r)e^{-i\phi}\end{pmatrix}.
		\label{solution}
\end{equation}
The spinor components $\chi_{\sigma}^{j,\nu}(r)$ can be given in closed form
for a step-like potential $U(r)=U_0 \theta (r-R)$ \cite{circular_QD}, 
however the eigenenergies $E_{n}$ have to be evaluated numerically. Each eigenstate is characterized by a pair $(n,s)$ where $s=\uparrow,\downarrow=\pm 1$ is the spin and where $n = (\nu,j)$ has a radial and angular momentum part.

\textit{In-Plane Phonons---}In order to study processes based on the admixture mechanism, we begin with the Hamiltonian $H=H_{0}+H_{\mm{SO}}+H_{\mm{EPC}}$, 
where $H_0$ describes the graphene QD without SOI as explained above,
$H_{\mm{SO}}$ describes the SOI,
and $H_{\mm{EPC}}$ describes EPC.
The effect of the SOI is to weakly mix the eigenstates Eq.~\eqref{solution}.
In this manner, e.g.\ the QD ground state, say $\ket{n=(0,1/2),\uparrow}^{(0)}$
acquires components of the excited states $\ket{n',\downarrow}^{(0)}$ 
with $n' =(\nu',j') \neq n$ and opposite spin, to first order in $H_{\mm{SO}}$,
\begin{equation}
	\ket{n\uparrow}=\ket{n\uparrow}^{(0)}+\sum_{n'\neq n}\frac{^{(0)}\bra{n'\downarrow\!}H_{\mm{SO}}\ket{n\uparrow}^{(0)}}{E_{n}-E_{n'}-\Ze}\ket{n'\downarrow}^{(0)},
	\label{perturbation}
\end{equation}
and similarly for $\ket{n\downarrow}$. 
With this admixed state the spin-conserving EPC can cause spin relaxation, 
\begin{multline}
	\bra{n\uparrow}H_{\mathrm{EPC}}\ket{n\downarrow}\equiv	          (H_{\mathrm{EPC}})_{nn}^{\uparrow\downarrow} = \\
\sum_{n' \neq n} \!\!\left[  \!	\frac{\left(H_{\mm{SO}}\right)_{nn'}^{\uparrow\downarrow}\left(H_{\mm{EPC}}\right)_{n'n}}{E_{n}-E_{n'}-\Ze} \! + \!	\frac{\left(H_{\mm{EPC}}\right)_{nn'}\left(H_{SO}\right)_{n'n}^{\uparrow\downarrow}}{E_{n}-E_{n'}+\Ze}
	\right]\!\!.
\label{eq:sum}
\end{multline}
For sufficiently small B-fields this can be expanded around $B=0$.
In the case of GaAs, 
the expression Eq.~(\ref{eq:sum}) vanishes  for $B=0$ due to the symmetry $\left(H_{\mm{SO}}\right)_{nm}^{\uparrow\downarrow}=-\left(H_{SO}\right)_{mn}^{\uparrow\downarrow}$ and $(H_{\mathrm{EPC}})_{nm}=(H_{\mathrm{EPC}})_{mn}$. 
This van Vleck cancelation \cite{van_vleck,nazarov} 
is one of the reasons for the high power of $B$ that appears in the spin
relaxation rate in GaAs QDs and can be traced back to the time-reversal
invariance of $H$ and its eigenstates, i.e., the fact that both SOI and 
EPC preserve time-reversal invariance.  
In particular, the spin relaxation takes
place from one state, say $\ket{n\!\!\uparrow}$, to its partner 
$\ket{n\!\!\downarrow}$ within a Kramers pair, which are linked by time reversal.

In our case, the states  $\ket{n\!\!\uparrow}$ and $\ket{n\!\!\downarrow}$ lie
in the same valley and therefore do not form a Kramers pair (see Fig.~\ref{fig:figure1}a).  The
time-reversed partner
of  $\ket{n\!\!\uparrow}$ is $\ket{n\!\!\downarrow}'$, where the prime 
denotes the opposite valley. 
Since neither the EPC nor the SOI lead to inter-valley 
mixing, spin-relaxation is effectively constrained to a single valley.
Therefore the selection of spin qubit states within the same valley
breaks time-reversal symmetry and leads to the absence of the
van Vleck cancelation.
We now proceed to the evaluation of the matrix elements of the SOI and
the EPC in Eq.~(\ref{eq:sum}) in order to calculate the spin relaxation rate.

We divide the SOI Hamiltonian into its intrinsic and Rahsba terms \cite{kane},  
\begin{equation}
	H_{\mm{SO}} =
	H_{\mm{i}}+H_{\mm{R}} = 	\Delta_{\mm{i}}\tau\sigma_{z}s_{z}+\Delta_{\mm{R}}(\tau\sigma_{x}s_{y}-\sigma_{y}s_{x}),
\label{SOI}
\end{equation}
where $\sigma_{i}$ and $s_{i}$ denote the Pauli matrices acting on the pseudo-spin and  real spin. 
We use a spin quantization axis aligned with the external B-field (see Fig.~\ref{fig:figure1}b)
and corresponding spinors $\ket{\!\!\uparrow_{B}}$ and
$\ket{\!\!\downarrow_{B}}$ and obtain
$f_x\equiv\bra{\uparrow_{B}\!\!}s_{x}\ket{\!\!\downarrow_{B}} = 
\cos^{2}\frac{\theta}{2}-e^{-2i\phi_B}\sin^{2}\frac{\theta}{2}$
and $f_y\equiv\bra{\uparrow_{B}\!\!}s_{y}\ket{\!\!\downarrow_{B}}
= -i( \cos^{2}\frac{\theta}{2}+e^{-2i\phi_B}\sin^{2}\frac{\theta}{2})$. 
First we consider $H_R$ and
calculate its matrix elements with states $\ket{n \uparrow_{B}}$ and $\ket{n' \downarrow_{B}}$. 
The two spin states we use are orthogonal, i.e. $\left<\uparrow_{B}|\downarrow_{B}\right>=0$ but they are not $s_{z}$ eigenstates.
In principle this allows both the intrinsic and Rashba SOI to provide a relaxation channel in the admixture mechanism.
However, due to the circular symmetry of the dot, selection rules for $j$ apply. 
In the case of $H_R$ only dipole transitions ($|j-j'|=1$) are allowed, whereas 
the matrix element of $H_i$ gives rise to selection rules $j=j'$ which turns out to be
incompatible with the selection rule $|j-j'|=1$ for the EPC.

The matrix element of $H_R$ can be written as
$\HRA = 2\pi \Delta_{R}\big[f^{y}(\dplus\NA+\dminus\NB) - i f^{x}(\dplus\NA-\dminus\NB)\big]$,
where $\NA=\int\mm{d}r\,r\,\chi^{n}_{A}\chi^{n'}_{B}$.
The matrix element $\HRA$ is neither symmetric nor antisymmetric
in contrast to the case of GaAs where an antisymmetry leads to van Vleck cancelation.

We consider two different EPC mechanisms which correspond to different changes in the lattice induced by phonons. The deformation potential  is caused by an area change of the unit cell, whereas the bond-length change mechanism corresponds to a modified hopping propability \cite{ando,mariani}. 
Because we work in the low-energy regime, we only consider acoustic phonons. 
In principle there are six possible relaxation channels:
(i) longitudinal acoustic (LA), transversal acoustic (TA), transversal out-of-plane (ZA) phonons, and 
(ii) deformation potential ($g_1$) and bond-length change ($g_2$) mechanisms.
In lowest order in the atomic displacement, the EPC has the form 
\cite{ando,mariani}
\begin{equation}
	H_{\mathrm{EPC}} = 
	\frac{q}{\sqrt{A \rho\omega_{\vec{q},\mu}}}
	\left(\begin{array}{cc}
		g_1 a_1 & g_2 a_{2}^{*} \\
		g_2 a_2 & g_1 a_1
	\end{array}\right) 
	\left(e^{i\vec{qr}} b^{\dagger} - e^{-i\vec{qr}} b\right),
	\label{EPC}
\end{equation}
with $a_1=i$ and $a_2=ie^{2i\phi_{\vec{q}}}$ for LA phonons, and $a_2=e^{2i\phi_{\vec{q}}}$ and $a_1=0$ for TA phonons, and $A$ the area of the graphene sheet. 
The vanishing of $a_1$ is due to the fact that in the regime of linear atomic displacements the coupling of the TA mode is a two-phonon process. 
Here, we restrict our considerations to one-phonon processes.
For a B-field of $B=1 \unit{T}$ and a sound velocity  of $s=2\times10^{4} \unit{m/s}$ \cite{falkovsky}, we obtain from $g\mu_{B}B=\hbar sq$ a phonon wavelength of $\lambda\approx 300 \unit{nm}$ which is an order of magnitude larger than a typical QD size of $25 \unit{nm}$ \cite{trauzettel}, thus
justifying the use of the dipole approximation for typical laboratory fields.

For the matrix element for LA phonon coupling via the deformation potential we find
$\left(H_{\mathrm{EPC}}^{\mathrm{LA}}\right)_{nn'} 
= -\frac{g_1 \pi}{\sqrt{A \rho s_{\mathrm{LA}}}} q^{3/2} M_{n n'} \left(\delta_{jj'+1}e^{-i\phi_{q}}+\delta_{jj'-1}e^{i\phi_{q}}\right)$
with
$M_{n n'}=\int\dr\; r^2 \left( {\chi_{A}^{n}}^* \chi_{A}^{n '}
+{\chi_{B}^{n}}^*\chi_{B}^{n'}\right)$.
The dependence on the phonon emission angle $\phi_{q}$ disappears upon summation over final states.
For the TA phonons we find that the coupling via the deformation potential is a two-phonon process which will not be discussed here.

The bond-length change mechanism leads to similar results for both LA and TA phonons,
$\left(H_{\mathrm{EPC}}\right)_{nn'} = D_i q^{1/2}
\left(\delta_{jj'+1}e^{-2i\phi_{q}}N^{AB}_{nn'}
\pm \delta_{jj'-1}e^{i2\phi_{q}}N^{AB}_{n'n}\right)$
with $D_{\mathrm{LA}}=-i2\pi g_2/\sqrt{ A \rho s_{\mathrm{LA}}}$ and 
$D_{\mathrm{TA}}=2\pi g_2 / \sqrt{ A \rho s_{\mathrm{TA}}}$, and
where the plus (minus) sign corresponds to LA (TA).
In linear order in the atomic displacement the ZA mode is decoupled from the other modes. The Hamiltonian Eq.~\eqref{EPC} cannot account for a coupling to the out-of-plane mode.

\begin{figure} 
	\includegraphics[width=0.48\textwidth]{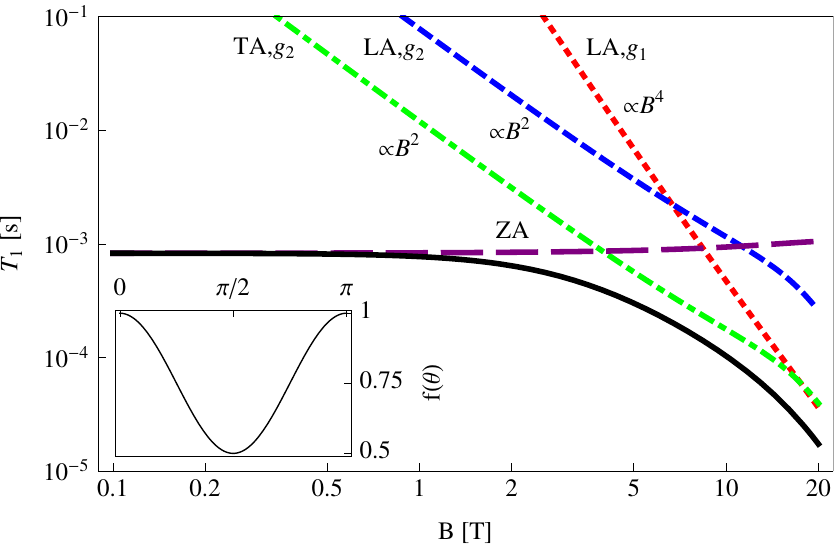}
	\caption{(color online) Log-log plot of the spin relaxation time $T_1$ as a function of an external B-field perpendicular to the plane ($\theta=0$) defined by the graphene sheet. The radius of the dot is $R=25 \unit{nm}$, both energy gap and depth of the dot are $260 \unit{meV}$. 
The individual relaxation channels are the coupling to LA in-plane phonons
via deformation potential ($g_1$, red dotted line) and the coupling to LA and TA phonons via bond-length change ($g_1$, blue dashed and green dot-dashed lines),
as well as the direct coupling to the out-of-plane (ZA) phonons (purple, long-dashed line). 
The black solid line represents the sum of all four processes.
	Inset:  Dependence of the relaxation rate on the inclination angle $\theta$ of the B-field.}
	\label{fig:figure2}
\end{figure}

With the matrix elements derived above,
we can write  the transition rates using Fermi's golden rule as
$1/T_1 \equiv \Gamma = 2\pi A\int\frac{\mathrm{d}^{2}q}{(2\pi)^{2}}\left|\left(H_{\mm{EPC}}\right)^{\uparrow\downarrow}_{nn}\right|^{2}\delta(sq-g\mu_{B}B)$.
For all mechanisms we find the same dependence on the orientation of the B-field, $f(\theta)=\cos^{4}(\theta/2)+\sin^{4}(\theta/2)=(3+\cos(2\theta))/4$. 
We find for the relaxation rate from the deformation potential
\begin{multline}
	\Gamma^{\mathrm{LA}}_{g_1} = 
	\frac{16 \pi^4 g_{1}^{2}\Delta_{R}^{2}}{\rho}
	\frac{(g\mu_{B}B)^{4}}{s_{LA}^{6}} f(\theta) \\
	\times\left|\sum_{n'\neq n}M_{nn'}R_{nn'}(\dplus\NA+\dminus\NB)\right|^{2},
\end{multline}
while for the bond-length change mechanism, we have
\begin{multline}
	\Gamma^{\mathrm{LA,TA}}_{g_2} = 
	\frac{64 \pi^4 g_{2}^{2}\Delta_{R}^{2}}{\rho}
	\frac{(g\mu_{B}B)^{2}}{s_{LA,TA}^{4}} f(\theta) \\
	\times\left|\sum_{n'\neq n}R_{nn'}
	\left(\dplus(\NA)^2 + \dminus(\NB)^2\right)
	\right|^{2},
\end{multline}
with  $R_{nn'}=(E_{n}-E_{n'})^{-1}$.
For numerical evaluation, we assume a QD size of $R=25\, \mathrm{nm}$ and 
$\Delta=10\delta$ where $\delta=v/R$ is the average level distance. 
The depth of the quantum well is also set to $U_0=10\delta$.
The Rashba SOI constant can be adjusted by an external electric field \cite{min} or by using different types of substrates. 
We chose a value of $\Delta_{R}=48 \unit{\mu eV}$ to calculate the relaxation times displayed in Fig.~\ref{fig:figure2}.
For the EPC constants we assume $g_1 = 30 \unit{eV}$ and $g_2 = 1.5 \unit{eV}$~\cite{ando}. 
We use as sound velocities $s_{\mm{LA}}=1.95\times 10^4 \unit{m/s}$ and $s_{\mm{TA}}=1.22\times 10^4 \unit{m/s}$~\cite{falkovsky}.
The overlap integrals $N^{AB}_{nn'}$ and $M_{nn'}$ are calculated numerically.  
The sum over $n'$ runs over all states, including the continuum. 
As shown in Table~\ref{tab:table1}, the contributions from higher levels vanish quickly so that we only take the first three levels into account. 
The relaxation rate $T_1=1/\Gamma$ is plotted in Fig.~\ref{fig:figure2}. 
\begin{table}
	\begin{tabular}{|c||c|c|c|} 
          \hline
		     & \bf{$\nu=1$} & \bf {$\nu=2$} & \bf{$\nu=3$} \\ 
		\hline \hline
		$j=-0.5$ & $1.1\times10^{4}$ & $2.6\times10^{-2}$ & $1.3\times10^{-3}$ \\ 
		$j=1.5$ & $1.6\times10^{4}$ & $1.2\times10^{1}$ & $9.3\times10^{-2}$ \\ 
		\hline
	\end{tabular}
	\caption{Individual relaxation rates in units of $s^{-1}$ for LA phonons via the deformation potential at $B=1\,{\rm T}$.  For higher quantum numbers $\nu$, the rate decreases quickly.}
	\label{tab:table1}
\end{table}

\textit{Direct Spin-Phonon Coupling---}
In flat graphene the acoustic phonons with perpendicular (ZA) polarization are decoupled from the in-plane modes (LA,TA).
We extend the SOI Hamiltonian Eq.~(\ref{SOI}) for the case of a  graphene layer
which is curved due to ZA phonons.
For displacements much smaller than the wavelength the normal vector of the graphene plane can be written as $\hat{\vec{n}}(z) \approx \hat{\vec{z}}+\nabla u_{z}(x,y)$ 
where $u_{z}(x,y)$ is the displacement field representing the ZA-phonons. 
Rotating the spin matrices into the local frame determined by the normal vector $\hat{\vec{n}}(z)$ 
we obtain in linear order in $u(z)$ a generalized SOI Hamiltonian 
$H_{\mm{SO}}=	H_{\mm{i}}+	H_{\mm{R}}$ with
\begin{eqnarray}
	H_{\mm{i}} =
	H_{\mm{i}}^{(0)}
	+ \Delta_{\mm{i}} 
	\left(\del_{x}u_{z}s_{x} + \del_{y}u_{z}s_{y} \right)
	\sigma_{z}\tau ,\label{SOI-curved-i}\\
	H_{\mm{R}} =
	H_{\mm{R}}^{(0)}
	+ \Delta_{\mm{R}} 
	\left(-\sigma_y \del_x u_z + \tau \sigma_x \del_y u_z\right) s_z ,\label{SOI-curved-R}
\end{eqnarray}
where $H_{\mm{i}}^{(0)}$ and $H_{\mm{R}}^{(0)}$ are the SOI Hamiltonians for flat graphene given in 
Eq.~(\ref{SOI}).
We evaluate these expressions for transverse out-of-plane (ZA) phonons,
with a quadratic dispersion relation $\omega_{\vec{q}}=\mu q^2$ where 
$\mu=\sqrt{\kappa / \rho}$ with $\kappa=1.1 \unit{eV}$
the bending rigidity and $\rho=7.5 \cdot 10^{-7}\unit{kg/m^2}$ 
the mass area density \cite{falkovsky,gazit}.
The EPC Hamiltonian is then obtained by substituting
the displacement operator for the ZA phonons 
$u_{z}=\sqrt{1 / A \rho \omega_{\vec{q}}}\left( e^{i\vec{q}\cdot\vec{r}} b^{\dagger} + e^{-i\vec{q}\cdot\vec{r}} b \right)$ into Eqs.~(\ref{SOI-curved-i}) and (\ref{SOI-curved-R}).
For the intrinsic SOI we obtain the matrix element 
	$(H_{\mm{i}})^{\uparrow\downarrow}_{nn} = 
	i\Delta_{\mm{i}} \sqrt{1/ A\rho \omega_{\vec{q}}} 
	\left(
	q_x \bra{\uparrow} s_x \ket{\downarrow} 
	+ q_y \bra{\uparrow} s_y \ket{\downarrow} 
	\right)
	\bra{n} \sigma_z e^{i\vec{q}\cdot\vec{r}} \ket{n} \nonumber$.
When evaluating the orbital matrix element only the lowest order in the dipole approximation contributes. All higher orders contain a factor $\propto e^{i\phi_{\vec{q}}}$ which averages to 
zero when the integration over $\phi_\vec{q}$ is carried out.

Finally, Fermi's Golden Rule is used to find the relaxation rate
\begin{equation}
	\Gamma^{\mm{ZA}} = \frac{2\pi^2 \Delta_{\mm{i}}^2}{\rho\mu^2} f(\theta)
	\left|
	\int\mm{d}r\; r \left(
	\left| \chi^{n}_{A}\right|^2 - \left| \chi^{n}_{B}\right|^2
	\right)
	\right|^2  , 
\end{equation}
which is  independent of $B$. 
The Matrix element itself depend only weakly on $B$. 
For the numerical evaluation we use $\Delta_{\mm{i}}=12 \unit{\mu eV}$~\cite{fabian} and $s_{\mm{ZA}}=1.59\times 10^3 \unit{m/s}$~\cite{falkovsky}.
The same calculation for the Rashba SOI yields vanishing matrix elements
and therefore no additional contribution.
In some cases, boundary conditions may lead to a linear dispersion relation for the ZA-phonons. We find that in this case the contribution due to ZA-phonons is negligible compared to the in-plane phonon contributions.

\textit{Conclusion---}We have calculated the electron spin relaxation time $T_1$ 
in a gate-tunable graphene QD arising from the combination of SOI and EPC.
We have restricted ourselves to the zero-temperature case, i.e. pure phonon emission which is
realistic at $0.1\,{\rm T}$ and $100\,{\rm mK}$ and higher temperatures for larger fields.
We have taken into account two mechanisms:  Admixture mechanism and direct spin-phonon coupling.
Due to selection rules in a circular QD, the admixture mechanism only leads to spin relaxation
in combination with the Rashba SOI.  The deformation potential 
EPC with LA phonons leads to a spin relaxation rate scaling as $B^4$ (Fig.~\ref{fig:figure2}), while
the bond length change EPC with both LA and TA phonons results in $B^2$ dependencies.
The relatively low powers compared to GaAs QDs can be traced back to the absence of the van Vleck cancelation, in combination with the 2D phonon density of states.
The direct coupling of electronic spins to ZA phonons only leads to spin relaxation in combination with
the intrinsic SOI whose rate does not depend on the applied B-field (in lowest order) and
thus leads to a B-field dependence at low fields which is markedly different from that in GaAs QD.

\textit{Acknowledgements---}We thank Andr\'as P\'alyi for useful discussions and 
we acknowledge funding from the DFG within FOR 912 ``Coherence and Relaxation Properties of Electron Spins''.

\end{document}